# Perfect Absorption Metasurfaces with Multiple Meta-Resonances

Suet To Tang, Joshua Lau, Ka Yan Au Yeung, Z. Yang*

Department of Physics, The Hong Kong University of Science and Technology, Clearwater Bay, Kowloon, Hong Kong, The People's Republic of China

**Abstract**

We show that the hybrid resonances of a DMR backed by a cavity are meta-resonances, in that they can be made as perfect as possible by fine tuning the structural parameters but without the requirements of extreme materials properties, such as zero dissipation. Instead, dissipation in the DMR is essential for the realization of perfect meta-resonances. We experimentally demonstrate such perfection by tuning the structure of a HMR till its reflection is as low as 0.426 % (1 – absorption = $1.8 \times 10^{-5}$). Besides the primary meta-resonances that are originated from the strong resonances of the DMR, weak hitchhiker resonances can also produce meta-resonances as perfect as the primary ones. The depth of the reflection dips is insensitive to the strength of the resonances involved, but critically depends on the degree of impedance match to air brought mostly by fine tuning the structure parameters, such as the cavity volume, the mass of the platelet, or the pre-tension in the membrane. Using the eccentricity of the platelet position in the DMR, a number of resonances and anti-resonances are generated, resulting in up to five meta-resonances within the range of 200 Hz to 1000 Hz, with the highest reflection being 7 % and the lowest being 1.2 % (1 – absorption = $1.4 \times 10^{-4}$). Other means of introducing hitchhiker meta-resonances are also reported.

# I    INTRODUCTION

The resonance of a text book classical harmonic oscillator is 'ideal' if the response function becomes infinitely large at its resonant frequency, but the presence of dissipation tames the infinity to a finite value, and spoils the otherwise perfect resonance. The maximum resonant response is inversely proportional to the dissipation coefficient. Therefore, in the presence of dissipation, the response can never be infinite. For anti-resonances of decorated membrane resonators (DMR's) the real part of the response function can be exactly zero [1]. Without the dissipation in the membrane, the transmission coefficient of the DMR will also be zero. The presence of dissipation leads to the non-zero imaginary part of the response function and non-zero transmission, spoiling the otherwise perfect sound blocking power of the DMR. The minimum transmission is proportional to the dissipation coefficient, which cannot be eliminated by structure designs. The minimum transmission of a DMR is therefore dissipation limited. Hybrid membrane resonators (HMR's), which is made by a DMR backed by a sealed gas cavity [2], along with several other types of functional structures [3 – 7], are termed 'perfect absorbers' because their absorption coefficient can reach above $\alpha = 0.99$, which is brought by near impedance matching of the device to air. However, absorption reaching 0.99 only requires the reflection to be less than 10%, even though in a recent work $\alpha = 0.999$ was accidentally achieved [6]. No closer attention has been paid to elucidate how close the absorption can be to the perfect value of unity, which can only be clearly revealed by pressing down the reflection as much as possible. In another work, the influence of the resistive layer in the device on the real part of the surface impedance was briefly studied [7], but no attention was paid to the imaginary part of the impedance which is also important for the device impedance to match to air. So far, all the works on absorption of functional structures seem to be content when $\alpha$ reached 0.99.

In the first part of the paper we show that the hybrid resonance of a HMR can be 'perfect' in that there is no limit on how small its reflection can be, even in the presence of intrinsic dissipation of the DMR. The presence of dissipation is in fact essential in achieving such perfect resonance. We experimentally demonstrate such perfection by tuning the structure of a HMR till its reflection is as low as 0.426 %, leaving only $1.8 \times 10^{-5}$ of the incident energy being reflection, or an absorption coefficient of $\alpha = 0.999982$. The hybrid resonance of a HMR therefore belongs to a different class of resonances in that it can be made to perfection by structural designs. We refer to such artificial perfect resonances meta-resonances, which are distinctly different from the 'natural' resonances, such as the harmonic oscillators and the DMR's, which can never be perfect in the presence of dissipation.

Acoustic metasurfaces with multiple frequency bands or even broadband perfect absorption are highly desirable for potential applications in many areas of noise abatement [5]. While multiple high absorption frequency bands have been realized with quite a few multiple-cavity devices made by HMR's and Helmholtz resonators with space coiling or spiral cavities [2 – 18], there is no report on multiple frequency perfect absorption by a single cavity, either in the form of HMR with one DMR or Helmholtz resonators. In the second part of the paper, we introduce the concept of hitchhiker meta-resonances for multiple frequency perfect absorption. In the impedance spectra, we observe that these resonances are 'riding' on the wake of the primary meta-resonances, and generate secondary (hitchhiker) meta-resonances with high degree of perfection (very low reflection). In one HMR with a single DMR and just one cavity, up to five perfect absorption peaks are experimentally observed within the 200 to 1000 Hz range, with the highest reflection being only 7 % ($\alpha$ = 0.9995) and the lowest being 1.2 % ($\alpha$ = 0.99986). Other means of introducing hitchhiker meta-resonances are also reported.

## II    EXPEIRMENTS

The structure parameters of the DMR's are summarized in Table 1. All the membranes are 0.02 mm thick. The circular mounts of the DMR's were made of 3D printing thermal plastics. The width and the thickness of the mounts are 5 mm. Photos of representative DMR's are shown in Fig. 1(a). The membranes of DMR-1 and DMR-2 are transparent polyvinyl chloride (PVC). They were intended to be made the same but the end products were slightly different. The membranes of the remaining DMR's are made of latex rubber. The decorating platelet is a steel washer 6 mm in diameter in all the DMR's or two stacked washers. DMR-3 and DMR-4 each consists of a DMR surrounded by large side orifices. The only difference between DMR-3 and DMR-4 is that the mass of the platelet of DMR-3 is twice of that of DMR-4. DMR-5 is similar to DMR-1 but with the platelet accidentally placed slightly off the center of the membrane. DMR-6 is a more exaggerated version of DMR-5 with the platelet placed well off center. Two types of measurements were performed. One is the transmission of a DMR, with the purpose of identifying its resonances and the anti-resonances [1] for later analysis of hybrid resonance when it is mounted on a sealed cavity to form a HMR. The other type of measurements is the reflection of the HMR mounted on a hard wall with negligible transmission.

All the transmission and the reflection spectra of the samples were measured using the standard impedance tube method. The details of the apparatus are given in our earlier work [2]. The transmission of each DMR was measured first. When measuring the transmission of DMR-3 and DMR-4, the side orifices were sealed off beforehand to ensure that the only path of sound transmission is through the DMR. Each DMR without side orifices was then mounted on a sealed hard plastic cavity to form a HMR, which was then mounted on a thick aluminum plate with a suitable opening while sealing off the remaining cross section area of the impedance tube.

| Sample | Membrane Diameter and Type | Decorating Mass (mg) | Side Orifices |
|--------|---------------------------|----------------------|---------------|
| DMR-1  | 60 mm PVC                 | 130                  | No            |
| DMR-2  | 60 mm PVC                 | 130                  | No            |
| DMR-3  | 40 mm Latex Rubber        | 260                  | Yes           |
| DMR-4  | 40 mm Latex Rubber        | 130                  | Yes           |
| DMR-5  | 40 mm Latex Rubber        | 130                  | No            |
| DMR-6  | 40 mm Latex Rubber        | 130                  | No            |

Table 1 List of the samples and their structural parameters.

## III    RESULTS AND ANALYSIS

### A. Highly perfect absorption

The transmission spectrum of DMR-1 is shown in Fig. 1(b) as the green curve. It exhibits a typical DMR transmission spectrum [1] with two resonant peaks near 317 Hz and 825 Hz, respectively, and an anti-resonance dip at 461 Hz. The reflection spectrum of the corresponding HMR-1 with DMR-1 backed by a cavity 30 mm in length and 60 mm in diameter (red curve) exhibits a dip at 368 Hz, which, as expected, is the hybrid resonance between the first resonance and the anti-resonance mediated by the cavity [2]. The results of HMR-1 resemble the HMR in our earlier work [2].

The hybrid resonance occurs whenever the impedance of air in the waveguide $Z_0$ matches the HMR impedance $Z_{HMR}$, i. e.,

$$\rho_0 c_0 S_{WG} = Z_0 = Z_{HMR} = S_M (Z_{DMR} - \frac{i\gamma P_0}{\omega V_C}) \qquad (1),$$

where $\rho_0$ is the air mass density, $c_0$ is the sound speed in air, $S_{WG}$ is the cross section area of the waveguide, $S_M$ is the membrane area, $Z_{DMR}$ is the impedance of the DMR, $\gamma$ is the adiabatic coefficient of air, $P_0$ is the static air pressure, $V_C$ is the cavity volume, $\omega$ is the angular frequency of the sound wave, and $i \equiv \sqrt{-1}$. The impedance of the cavity is $Z_C = -\frac{i\gamma P_0 S_M}{\omega V_C}$ [2]. The impedance of the DMR is given by [19]

$$\frac{1}{Z_{DMR}} = \sum_n \frac{i\omega |<W_n>|^2}{m_n(\omega_n^2 - \omega^2 + i\omega\beta_n)} \quad (2)$$

where $W_n$ is the displacement field of the *n-th* resonant mode, the '< >' sign stands for taking average over the DMR surface, $m_n = \frac{1}{S_M}\int \rho |W_n|^2 d\Omega$ is the displacement-weighted mass, $\rho$ is the mass density distribution of the decorated membrane, $\omega_n$ and $\beta_n$ are the angular frequency and the dissipation coefficients of the *n-th* resonant mode of the DMR. For the resonant modes with perfect azimuth anti-symmetry vibration pattern, $<W_n> = 0$, so they do not contribute to the DMR impedance.

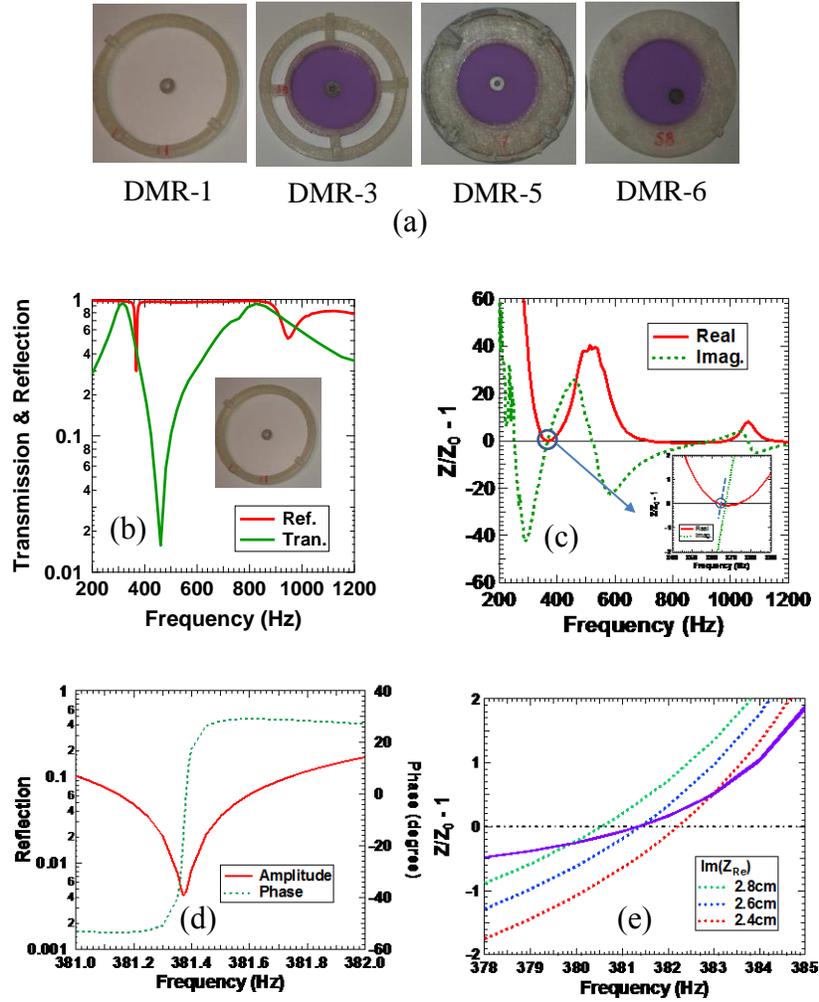

Figure 1 (a) Photos of the representative DMR's used in this work; (b) The transmission spectrum of DMR-1 (green curve) and the reflection spectrum (red curve) of the corresponding HMR-1; (c) The reduced surface impedance of HMR-1. The insert is a magnified region of the spectra where the HMR-1 impedance is close to that of air. (d) The reflection amplitude (solid red curve) and phase (dashed green) spectra of HMR-2; (e) The reduced impedance of HMR-2 at several cavity lengths. The real part of the impedance is represented by the solid purple curve as it is independent of the cavity length.

The experimental real and the imaginary parts of the reduced impedance $Z_{Re} \equiv \frac{Z_{HMR}}{Z_0} - 1 = \frac{2R}{1-R}$ shown in Fig. 1(c), where $R$ is the complex reflection coefficient, reveal more clearly what happens. When both the real and the imaginary parts of $Z_{Re}$ are zero at the same frequency, perfect hybrid resonance and unity absorption occurs. Near the hybrid resonance, the real part of the impedance Re($Z_{HMR}$) drops drastically from much larger than that of air $Z_0$ to even below $Z_0$, while Im($Z_{HMR}$) passes through the zero line (See the insert of Fig. 1(c)). Without the cavity (zero cavity volume) $Z_C$ will be infinitely large and negative while purely imaginary. Then Im($Z_{Re}$) would never cross the zero line so no hybrid resonance could ever occur. This is the scenario of a DMR backed by a hard wall without any spacing in between. In the case shown in the insert of Fig. 1(c), Re($Z_{Re}$) happens to cross the impedance matching line at two frequencies, one of which is marked by the circle. The resulting reflection dip is not very small (~ 0.3, $\alpha$ = 0.91) because the crossing frequencies of the real and the imaginary parts of $Z_{Re}$ differ by about 3 Hz. If Im($Z_{Re}$) could be upper lifted by about 1.4 as represented fictively by the short dashed line, it would cross the zero line at the same frequency where Re($Z_{Re}$) = 0, i. e., achieving perfect impedance matching. This could be realized by fine-tuning the cavity volume, for example. According to Eq. (1), changing the cavity volume would only change the imaginary part of $Z_{Re}$. Given the structure parameters of HMR-1, changing Im($Z_{Re}$) by 1.4 would require the increasing of cavity length by about 1 cm and the fine tuning precision below 1 mm.

To test the assumption to realize perfect hybrid resonance by tuning the cavity volume, we fabricated HMR-2 with DMR-2 mounted on a cavity 60 mm in inner diameter, the length of which could be fine-tuned in the vicinity of 30 mm by a micrometer. The results in Fig. 1(d) show that when the cavity length was changed from 26 mm to 28 mm, Re($Z_{Re}$) (the purple curves which cannot be resolved) remains almost the same, while Im($Z_{Re}$) at different cavity length form a family of curves. This is consistent with Eq. (1) in that only Im($Z_{Re}$) is dependent on the cavity volume. By fine tuning the cavity length the lowest reflection obtained is 0.426 %, and less than 1.8 × 10$^{-5}$ part of the incident wave energy is reflected. The relative mismatch to air impedance, given by $\frac{Z_{HMR} - Z_0}{Z_0} = 2R$ for $R \ll 1$, is less than 0.86 %. Also note that the membrane area is 3.5 times smaller than that of the waveguide.

The spectral feature in Fig. 1(d) has all the characteristics of a resonance, i. e., well defined and narrow line shape in amplitude accompanied by a sudden change in the phase. Besides the obvious benefit of high absorption in the low frequency range by a meta-surface made by a 2D array of such HMR's, the extremely narrow line width of the reflection dip could provide a pathway for realizing extremely narrow notch filters and phase sensitive detectors. In this particular case, near the meta-resonance frequency a change of 0.01 Hz in frequency can change the reflection relative value by 20 %, 0.02 Hz to nearly double the reflection value, and 0.015 Hz to shift the phase by 26 degrees.

In mechanical vibrations the presence of dissipation usually spoils the otherwise perfect resonances. For an elastic body, its resonance is 'perfect' when the response function becomes infinity at a resonance frequency, but the presence of dissipation tames the infinity to a finite value. For anti-resonances of a DMR the real part of the response function can be exactly zero. The presence of dissipation in the DMR leads to the non-zero imaginary part of the response function and non-zero transmission, spoiling the otherwise perfect sound blocking capability of the DMR. In coherent perfect channeling [20], there is always some energy loss in the channeling process by the internal dissipation of the scatterer. Therefore, the presence of dissipation will spoil the perfection of the channeling process. The hybrid resonance, on the other hand, belongs to another class of resonances, which we refer to as meta-resonances, in that they can be made to perfection by structure designs even in the presence of dissipation in the structural components of the devices. The meta-resonance is realized by changing the structure, rather than relying on extreme materials properties such as zero dissipation. In fact, HMR meta-resonances are only possible when dissipation is present in the DMR. Without dissipation in the DMR, $Z_{HMR}$ would be purely imaginary so perfect matching condition $Z_{Re} = 0$ could never be reached. In the present example, the loss factor of the membranes is of the order of 1 %. If the loss factor is too large, the resonance strength will not be strong enough to pull down $\text{Re}(Z_{Re})$ to reach the zero line.

### B. Multiple band perfect absorption by hitchhikers

Realizing multiple frequency hybrid resonance (meta-resonance) with a single cavity could significantly broaden the means of noise abatement, in addition to deepening the understanding of the hybrid resonances. HMR-1 is a 'near ideal' structure, as compared to the ones presented below, in that the reflection dip only occurs at the frequencies between a primary resonance and an anti-resonance of an ideal DMR. For example, the second reflection dip of HMR-1 is at 946 Hz with an unimpressive minimum value of 0.52. Numerical simulations show that a second anti-resonance is beyond the experimental limit of 1500 Hz, and this reflection dip is of the same type as the first one, which we refer to as primary meta-resonance. The impedance spectra in Fig. 1(c) show that near 950 Hz Im($Z_{Re}$) crosses the zero line but Re($Z_{Re}$) is not small enough, i. e., the additional impedance due to the cavity is not well suited for perfect meta-resonance near that frequency. If the cavity length is adjusted to improve the impedance matching there, the impedance matching at the first reflection dip at 368 Hz would be compromised. This is the situation often encountered when one attempts to realize multiple frequency perfect absorption with a single cavity.

The difficulty can be overcome by introducing a resonance of weaker strength to 'hitchhike' on the wake of the primary hybrid resonance as exemplified by HMR-1. In particular, we placed DMR-3 at 12 mm in front of HMR-1. It can be seen from the transmission of DMR-3 (blue curve in Fig. 2(a)) that its 1$^{st}$ and 2$^{nd}$ resonances are at 160.7 Hz and 677.8 Hz, respectively, and its anti-resonance is at 237.7 Hz. The reflection spectrum of the combined hybrid resonance device (red curve) along with the transmission of DMR-1 (green curve) is shown in Fig. 2(a). Due to the large orifices of Sample-3, HMR-1 remains largely unchanged, and its reflection dip is at nearly the same frequency as the original HMR-1. Another reflection dip appears at 654 Hz, obviously due to the presence of DMR-3. However, rather than occurring somewhere between the 1$^{st}$ resonance and the anti-resonance of DMR-3, this hybrid resonance is above the anti-resonance and close to but below the 2$^{nd}$ resonance of DMR-3. Compared to the $Z_{Re}$ in Fig. 1(c), the $Z_{Re}$ of the combined hybrid resonance device has an additional resonant feature originated from the second resonance of DMR-3. Such resonance brings Im($Z_{Re}$) to cross the zero line once more than the original HMR-1. When the spacing between DMR-3 and HMR-1 is appropriately adjusted, which shifts mostly the imaginary part of the impedance of DMR-3 but not that of HMR-1, a nearly perfect meta-resonance occurs, with minimum reflection of only 7 %. If the position of DMR-3 is further fine-tuned, it is possible to realize a meta-resonance with perfection comparable to that of HMR-2.

To further verify that the new reflection dip is independent of the anti-resonance of the hitchhiking DMR, we replaced DMR-3 with DMR-4 with half the platelet mass of DMR-3. As is well known [1], the change of the platelet mass would change the 1$^{st}$ resonance and therefore the anti-resonance frequency, but hardly the 2$^{nd}$ resonance frequency. For DMR-4 the 1$^{st}$ resonance is at 209.46 Hz. Its anti-resonance is at 328.16 Hz. Both are significantly higher than the corresponding ones of DMR-3, as expected. Its 2$^{nd}$ resonance is at virtually the same frequency of 678 Hz as that of DMR-3. As expected, the reflection dip due to DMR-4 at 12 mm from HMR-1 is at 645.2 Hz, close to that of DMR-3 at 654 Hz, with a minimum value of 15 %. The spectra are not shown due to space limit.

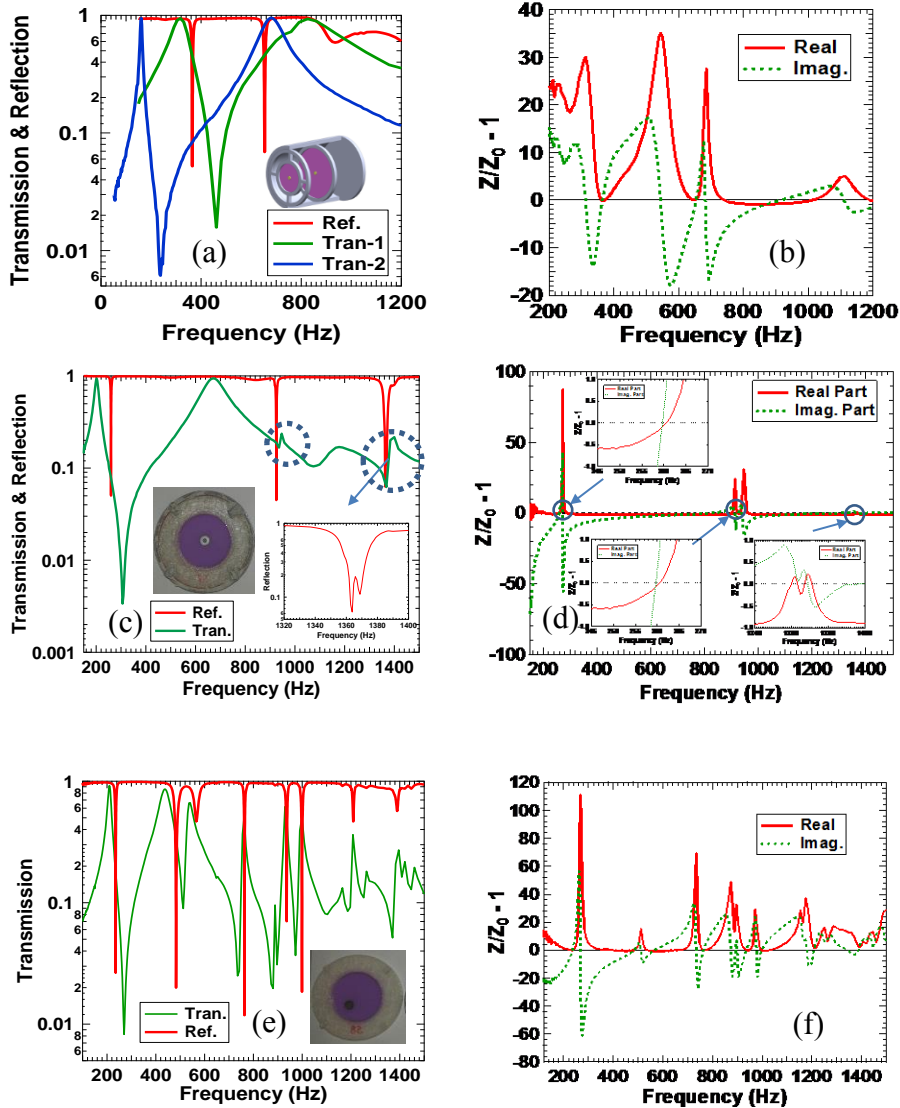

Figure 2 (a) The transmission spectrum of DMR-3 (blue curve), the transmission spectrum of DMR-1 (green curve), and the reflection spectrum of the combined sample (red curve). Its structure is schematically shown in the insert; (b) The reduced impedance of the combined sample in (a); (c) The transmission spectrum of DMR-5 (green curve) and the reflection spectrum of the corresponding HMR-5 (red curve). The right hand side insert is a magnified view of the double-dip feature in the reflection; (d) The corresponding reduced impedance of HMR-5. The inserts depict the magnified views of the frequency regions where near impedance matching occurs; (e) The transmission spectrum of DMR-6 (green curve) and the reflection spectrum of the corresponding HMR-6 (red curve). The insert is a photo of DMR-6; (f) The reduced impedance of HMR-6.

The above results and analysis indicate that when the resonances are well separated in frequency, the total impedance of the combined structure of a hitchhiker DMR-3 in front of a host HMR-1 could be viewed as a simple superposition of the individual structures to the zeroth order approximation. The presence of DMR-3 or DMR-4 also perturbs HMR-1 in a small but positive way, in that its reflection dip frequency (364.4 Hz) is shifted a little bit from the original 368 Hz and its minimum value is reduced to 5.2 %, which is about 6 times smaller than the original HMR-1. Such improvement, however, is accidental, as the detailed perturbation could only be analyzed by careful numerical simulations, which is outside the scope of this paper.

The above case is an example of hitchhiker meta-resonances. When riding on the wake of the primary (and strong) resonances of the host HMR, the hitchhiker resonance only needs to bring down the real part of the combined impedance by a small amount to reach the zero line, while introducing a zero-line crossing in the imaginary part of the impedance. When the imaginary part is tuned properly by adjusting the position of the hitchhiker DMR relative to the host HMR, a meta-resonance occurs. These hitchhiker meta-resonances do not require the participation of the anti-resonance of the hitchhiker DMR. The presence of the anti-resonance of DMR-1 in the primary HMR-1 provides the necessary low impedance background in the relevant frequency range, in which a hitchhiker resonance, such as the one brought by DMR-3 or DMR-4, could produce a highly perfect meta-resonance.

### C. Multiple-band perfect absorption by DMR eccentricity

DMR's like DMR-1 with perfect circular symmetry only have primary resonances with perfect circular symmetry vibration patterns [1]. They are well separated, and there are large frequency ranges in between where the impedance is still quite small, owing to the tails of the strong resonance strength of the primary resonances. Following the same strategy of introducing secondary resonances to produce hitchhiker meta-resonances, we fabricated DMR-5 and the corresponding HMR-5. The optimum cavity length is 67 mm and its cross section is the same as the DMR. The transmission spectrum of DMR-5 (the green curve in Fig. 2(c)) indicates that its $1^{st}$ and $2^{nd}$ primary resonances are at 200 Hz and 670 Hz, respectively. The anti-resonance is at 306.3 Hz. Besides these prominent features, there are two weak wiggling features marked by the circles, one around 900 Hz and the other around 1300 Hz. From the amplitudes of the features, they can be identified and confirmed by simulations as weak secondary resonances due to small eccentricity of the platelet position. Without the eccentricity the vibration patterns of the corresponding resonances would have perfect azimuth anti-symmetry, so they cannot be excited by the uniform incident sound waves. In the presence of the structural eccentricity, these vibration patterns no longer possess perfect azimuth anti-symmetry, so they can be weakly excited by the sound waves. In terms of the expression given by Eq. (2), the areal average of the vibration field is small but no longer exactly zero. The reflection spectrum of HMR-5 exhibits three dips. The one at 259.7 Hz with minimum value of 5.0 % can be identified as the primary meta-resonance as in HMR-1. The ones at 925 Hz and around 1365 Hz, however, are rather unexpected, because the transmission spectrum indicates that these are weak resonances. The impedance spectra in Fig. 2(d) reveal the origin of these meta-resonances. Riding on the background of small impedance region produced by the primary meta-resonance at 259.7 Hz, these weak resonances produce hitchhiker meta-resonances. The strength of the resonant feature around 900 Hz is about 1/3 of the primary one around 300 Hz. Its meta-resonance is at 925 Hz with a minimum reflection of 4.5 %. The astonishingly weak resonance around 1360 Hz in fact produces two reflection dips with minimum values of 6.3 % at 1364 Hz and 11 % at 1369 Hz. The resonance strength is as small as that of air $Z_0$, in stark contrast to the ~ $90Z_0$ in strength generated by the primary resonance near the first meta-resonance. It serves as a vivid example that the perfection of the meta-resonances is independent of the resonance strength they are originated from.

To further explore the benefit of eccentricity we fabricated DMR-6 with the platelet at 10 mm off the center of the membrane. The cross section of the cavity for the corresponding HMR-6 is the same as HMR-5, and its optimum length is 40 mm. As shown in Fig. 2(e), the first transmission peak and dip of DMR-6 is about the same as DMR-5. The first reflection dip of HMR-6 at 235.5 Hz is comparable to that of HMR-5 at 259.7 Hz. Corresponding to the second transmission peak of DMR-5 are two transmission peaks in DMR-6, which produce another fairly strong anti-resonance in between. In between the anti-resonance and the $2^{nd}$ resonance is the second reflection dip at 483.2 Hz. The small wiggle in DMR-5 around 900 Hz is now amplified by the large eccentricity of the platelet position, resulting in several transmission peaks and dips of fairly large strength. Between these peaks and dips are three reflection dips at 763.6 Hz, 936.2 Hz, and 998.9 Hz, with minimum reflection of 1.2 %, 7.0 %, and 1.8 %. The impedance spectra in Fig. 2(f) show that the Im($Z_{Re}$) crosses the zero line six times and produces five deep and one shallow reflection dips. As the vibration patterns of the resonances of DMR-6 no longer have clear azimuth symmetry or anti-symmetry, the incident sound waves can excite all of them with comparable strength. It is therefore difficult to identify which are primary ones and which are hitchhikers. Instead, it is a clear example that many meta-resonances with very high degree of perfection can be produced by a DMR with closely packed resonances and anti-resonances backed by a single cavity. The minimum reflection of these meta-resonances can be tuned by adjusting the cavity volume, so the strength of the resonances they originated from does not play any essential role in determining their perfection (minimum reflection values).

Simulations show that at hybrid resonances and membrane structure vibrate strongly to dissipate the incident energy, even though some resonances are apparently weak in the impedance spectra because the areal average of the membrane vibration field is small due to nearly equal in-phase and opposite-phase vibration field. In other words, vibrational energy can be strongly dissipated when the amplitude of the vibration patterns is strong but the areal average is near zero.

## IV    SUMMARY

In summary, the hybrid resonances of a DMR backed by a cavity are meta-resonances, in that they can be made as perfect as possible by fine tuning the structural parameters. Dissipation in the DMR is essential for meta-resonances. Besides primary meta-resonances which are originated from the strong resonances of the DMR, weaker or hitchhiker resonances can also produce meta-resonances as perfect as the primary ones. The depth of the reflection dips is insensitive to the strength of the resonances involved, but critically depends on the degree of impedance match to air brought mostly by fine tuning the structure parameters, such as the cavity volume, the mass of the platelet, or the pre-tension in the membrane. Using eccentricity of the position of the platelet in the DMR backed by a single cavity, a number of resonances and anti-resonances are generated, resulting in up to five meta-resonances within the range of 200 Hz to 1000 Hz. From the width of the $1^{st}$ resonance in the transmission spectra one can tell that the dissipation or loss factor of the PC membranes in DMR-1 and DMR-2 is larger than that of the latex membranes in the other DMR's. Yet both types of DMR's can produce meta-resonances with high perfection, indicating that structure designs are far more important than material properties, which is the most important advantage of the meta-materials in general over conventional materials.

### Acknowledgement

This work was supported by AoE/P-02/12 from the Research Grant Council of the Hong Kong SAR government.